# A room temperature polar ferromagnetic metal

Hongrui Zhang[1,10*], Yu-Tsun Shao[2,10], Rui Chen[1,3,10], Xiang Chen[3,4,10], Sandhya Susarla[1,3], Jonathan T. Reichanadter[5,6], Lucas Caretta[1], Xiaoxi Huang[1], Nicholas S. Settineri[6], Zhen Chen[2], Jingcheng Zhou[1], Edith Bourret-Courchesne[3], Peter Ercius[7], Jie Yao[1,3], Jeffrey B. Neaton[3,4,8], David A. Muller[2,9], Robert J. Birgeneau[3,4], Ramamoorthy Ramesh[1,3,4*]

[1] Department of Materials Science and Engineering, University of California, Berkeley, CA, USA.

[2] School of Applied and Engineering Physics, Cornell University, Ithaca, NY, USA.

[3] Materials Sciences Division, Lawrence Berkeley National Lab, Berkeley, CA, USA.

[4] Department of Physics, University of California, Berkeley, CA, USA.

[5] Department of Electrical Engineering, University of California, Berkeley, CA, USA.

[6] Department of Chemistry, University of California Berkeley, Berkeley, CA, USA.

[7] The Molecular Foundry, Lawrence Berkeley National Laboratory, Berkeley, CA, USA.

[8] Kavli Energy Nanosciences Institute at Berkeley, Berkeley, CA, USA

[9] Kavli Institute at Cornell for Nanoscale Science, Cornell University, Ithaca, NY, USA.

[10] These authors contributed equally.

*Corresponding authors. Email: hongruizhang@berkeley.edu, rramesh@berkeley.edu

The advent of long-range magnetic order in non-centrosymmetric compounds has stimulated interest in the possibility of exotic spin transport phenomena[1,2,3,4,5,6] and topologically protected spin textures[7,8,9,10] for applications in next-generation spintronics. This work reports a novel wurtzite-structure polar magnetic metal, identified as AA'-stacked $(Fe_{0.5}Co_{0.5})_{5-x}GeTe_2$, which exhibits a Néel-type skyrmion lattice as well as a Rashba-Edelstein effect at room temperature. Atomic resolution imaging of the structure reveals a structural transition as a function of Co-substitution, leading to the polar phase at 50% Co. This discovery reveals an unprecedented layered polar magnetic system for investigating intriguing spin topologies and ushers in a promising new framework for spintronics.

Non-centrosymmetric magnets offer an extraordinary platform for exploring fascinating magnetic, quantum topological phases, owing to broken crystal symmetries. As an example, magnetic skyrmions[11,12] and spin Hall effects[5,6,13] are observed in polar magnets and are usually under-pinned by an antisymmetric Dzyaloshinskii-Moriya interaction (DMI). In polar magnets such as $GaV_4S_8$ ($C_{3v}$)[7], $VOSe_2O_5$ ($C_{4v}$)[8], $GaV_4Se_8$ ($C_{3v}$)[9], and PtMnGa ($C_{3v}$)[10] with $C_{nv}$ crystal symmetry, the DMI confines the magnetic modulation direction vector perpendicular to the polar axis. Thus, a Néel-type skyrmion lattice is stable once a suitable magnetic field is applied along the polar axis. In addition, in polar metals or semiconductors[6,14], the crystal structure leads to a built-in electric potential along the polar axis, and spin orbit interaction is at the essence of the Rashba effect[15]. The Rashba Hamiltonian can be expressed as: $H_R = \alpha_R(\vec{k} \times \vec{\sigma}) \cdot \vec{z}$, where $\alpha_R$ is the Rashba coefficient, $\vec{k}$ is the momentum vector, $\vec{\sigma}$ is the Pauli matrix vector, and $\vec{z}$ is the unit vector along with the polar axis. In such a system, the Edelstein effect[16] is also observed, that is, a nonequilibrium spin accumulation occurs by applying an electric field to the spin-polarized bands. Thus, a highly efficient spin-to-charge conversion has been demonstrated in polar magnetic semiconductors[6] and interfacial systems[5,17,18,19]. Until now, few polar magnetic metals have been observed at room temperature. However, in Van der Waals(vdW)systems, the stacking configuration can be engineered to manipulate crystal symmetry, for example by synthesis conditions[20], chemical doping[21] and external field[22], and thus plays a crucial role in mediating physical phenomena such as ferroelectricity,[23] magnetism,[24] and superconductivity[25].

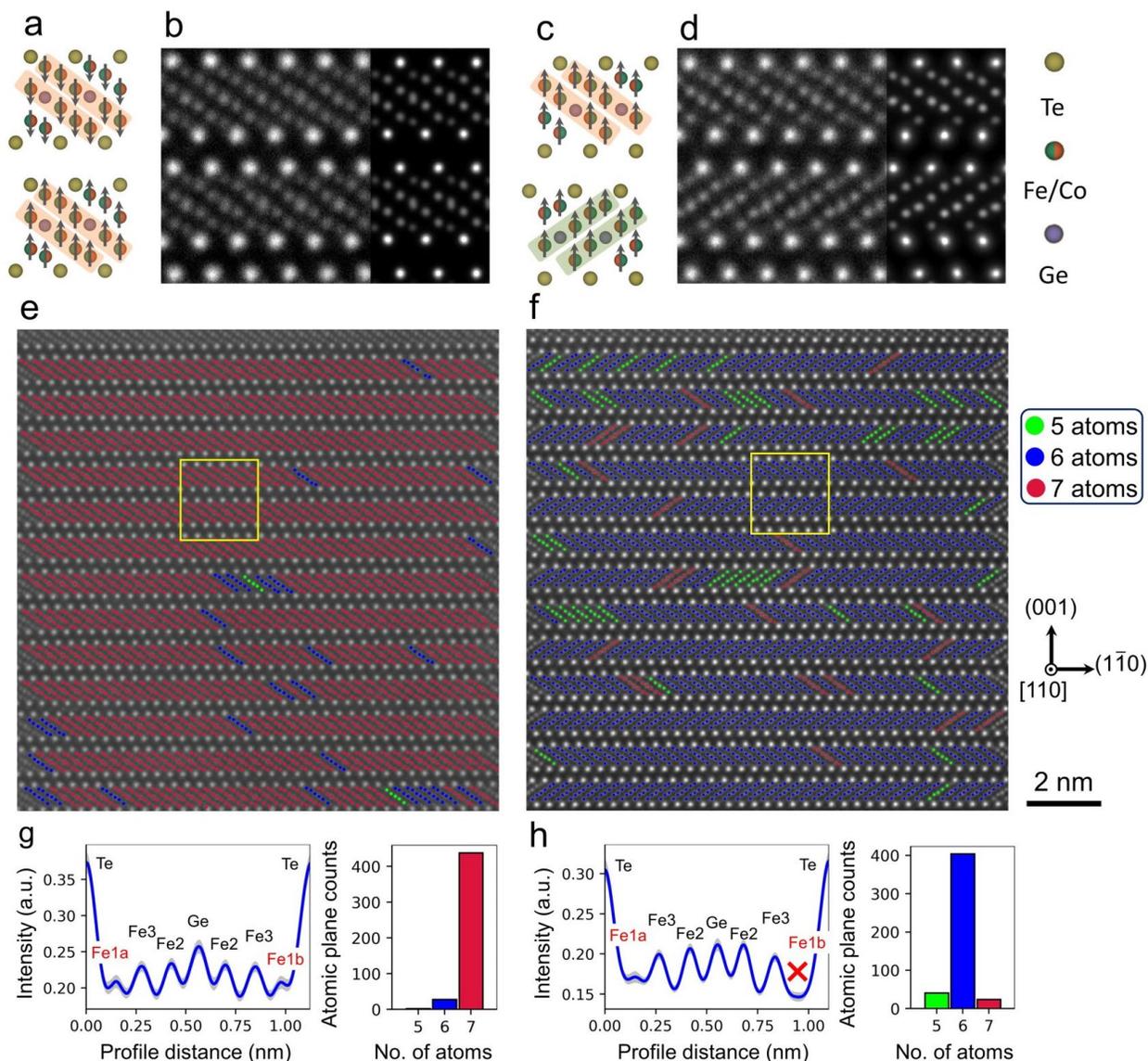

**Fig. 1: Structural phase transition of $(Fe_{1-y}Co_y)_5GeTe_2$.** **a,b,c,d,** Structure model and corresponding HAADF-STEM cross-sectional images of FCGT for (**a,b**) AA stacking (y=0.45) and (**c,d**) AA' stacking (y=0.50) along the [110] direction. Enlarged experimental and corresponding simulated HAADF-STEM images of a 10-nm-thick FCGT for AA-stacking (**b**) and (**d**) AA'-stacking phases. **e,f,** The Fe-Co-Ge-sub-lattices are color-coded by the number of atoms in the atomic planes, with 5(G), 6(B), 7(R) atoms, respectively. The yellow boxes in (**e, f**) indicate the regions for (**b, d**), respectively. **g,h,** Te-Te intensity line profiles and the histograms show the counts for number of atoms in each sub-lattice plane. The gray shaded area indicates the standard deviation of the intensity profiles.

The Fe$_N$GeTe$_2$ (N = 3,4,5) system exhibits a two-dimensional itinerant ferromagnetism and a high Curie temperature which provides an ideal platform to design such a metallic, polar magnetic material. Recent work reported the synthesis of a vdW Fe$_{5-x}$GeTe$_2$ magnet exhibiting rhombohedral ABC stacking with a ferromagnetic ground state.[26,27] Interestingly, the introduction of cobalt into the iron sites within this compound has been shown to induce a structural phase transition to the inversion-centric, hexagonal AA stacked (Fe$_{1-y}$,Co$_y$)$_{5-x}$GeTe$_2$ (y = 0.44, 0.46) with an antiferromagnetic ordering.[28,29] Here, we report the discovery and synthesis of a hexagonally-stacked (Fe$_{0.5}$Co$_{0.5}$)$_{5-x}$GeTe$_2$ (FCGT) phase, denoted as AA'- stacked FCGT, by systematically increasing the cobalt doping to 50%. Distinct from the previously identified Fe$_N$GeTe$_2$ systems, the AA' stacked FCGT exhibits a non-centrosymmetric crystal structure that shows ferromagnetic order with a remarkably high Curie temperature ($T_c$, ~365 K). As a result of this non-centrosymmetric structure, we observe a well-ordered lattice of Néel-type skyrmions at room temperature probed directly via Lorentz scanning transmission electron microscopy (L-STEM), magnetotransport measurements and magnetic force microscopy (MFM). Spin torque ferromagnetic resonance (ST-FMR) studies show a strong Rashba-Edelstein effect, illustrating the potential for use as both the ferromagnet and a spin orbit torque metal in spintronic applications.

High-quality FCGT single crystals with the two different stacking symmetries were synthesized by tuning the cobalt concentration using a chemical vapor transport method (details of the synthesis are presented in the Methods section). The composition of the FCGT platelets was confirmed by energy dispersive X-ray spectroscopy (EDS), illustrated for the 45% Co and 50% Co platelets (Fig. S1). Armed with the chemical stoichiometry, we then studied the atomic scale structure of these two platelets. Pristine Fe$_5$GeTe$_2$ has an ABC stacking with a rhombohedral unit cell and space group $R\bar{3}m$ (No. 166);[28,29] introduction of cobalt of upto 40-47% atomic concentration transforms the FCGT crystal into the AA phase with the space group $P\bar{3}m1$ (No. 164).[28] Surprisingly, as the cobalt concentration is increased to 50%, the AA stacked FCGT transitions into the AA' stacking which belongs to the space group $P6_3mc$ (No. 186) (see Methods and single crystal X-ray diffraction data in Fig. S2 and Fig. S3). To understand the origins of inversion symmetry in the AA and AA' phases at the atomic level, we performed scanning transmission electron microscopy (STEM) studies. Fig. 1 illustrates the atomic configuration for the two types of FCGT, labelled as AA (Fig. 1**a**) and AA' stacking (Fig. 1**c**). High angle annular dark field-scanning transmission electron microscopy (HAADF-STEM) images of cross-sections in the [110] zone axis of the two platelets reveal strikingly different stacking information, shown in Fig. 1(**b, d**) and Fig. S4 and Fig.S5. The corresponding simulated images are shown to the right of the actual images. The averaged intensity line traces for both these images reveal a significant difference, as shown in Fig.1(**g, h**). Both phases are built up of double Te layers between which the Fe/Co/Ge layers are stacked. The AA-phase (45% Co) has a predominance of 7-layer stacking with a minority of 6-layer stacking. In stark contrast, the AA' structure in the 50% Co sample has a predominance of 6-layer stacks with a minority of 5- and 7-layer stacks. It is also noteworthy that the three different Fe-sites are not at the same intensity, hinting at the possibility of partial occupancy of these sites, particularly the Fe1a and Fe1b sites. This difference in stacking sequence between the two Te layers is captured at a larger scale in Fig.1(**e, f**) for the AA and the AA' phases respectively. These images are false-color coded to illustrate the differences in stacking sequence, that was derived from atomic columns that were Gaussian-fitted to extract their exact positions. The bar charts to the right of Fig.1(**g, h**) summarize the statistics of the layer stacking for the AA and AA' phases.

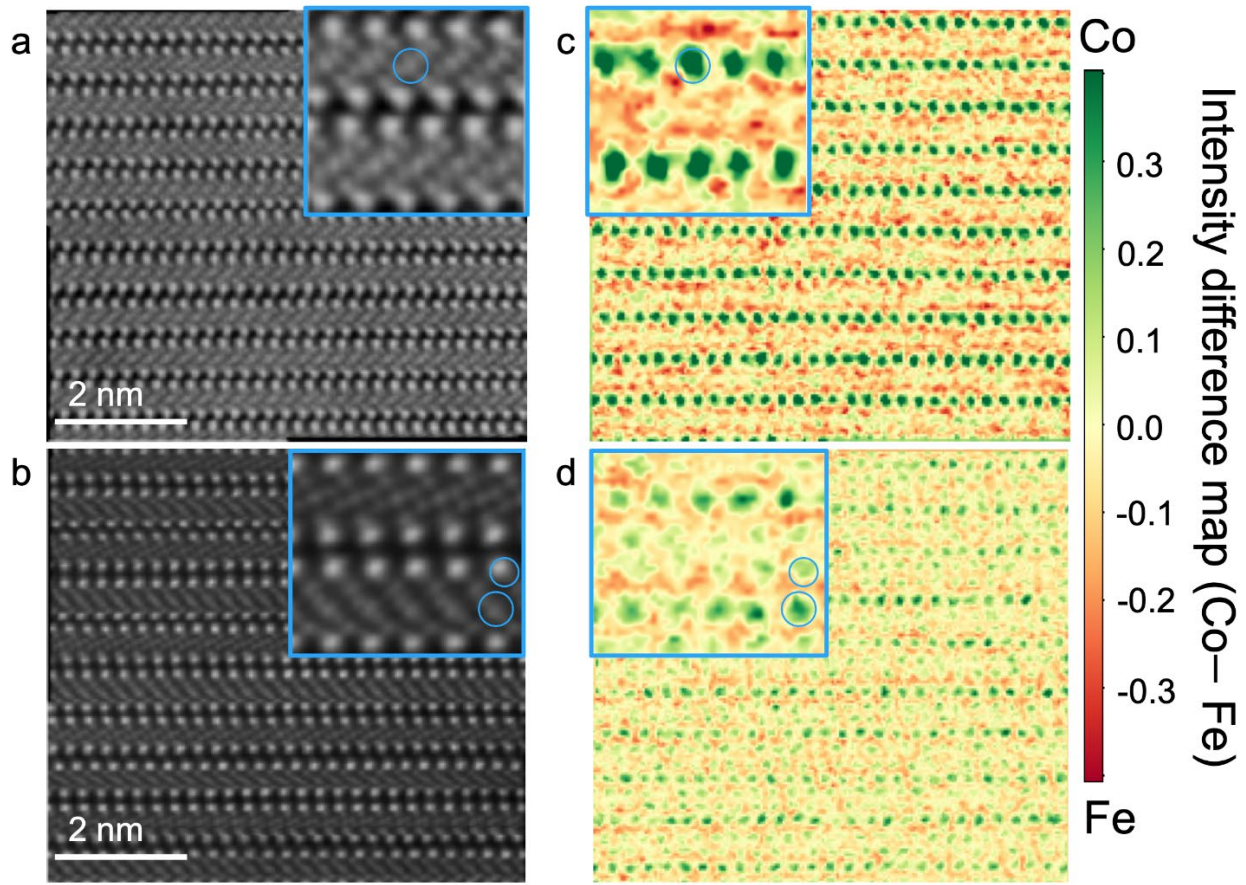

**Fig. 2: Chemical distribution in CFGT. a, b,** Simultaneous atomic resolution HAADF-STEM images for AA and AA' phase. **c, d,** Intensity difference EELS maps (Co - Fe) for AA and AA' phase. Inset indicates the zoomed-in images of the HAADF and EELS maps. Green coloring in the STEM-EELS maps indicates relatively higher cobalt concentration, and orange coloring indicates relatively higher iron concentration. The blue circles in the insets indicate the estimated delocalization radius of the EEL signal.

Armed with this atomic scale structural information, we then performed chemical mapping via simultaneous HAADF-STEM and electron energy loss spectroscopy (EELS). Details of the STEM-EELS acquisition parameters are described in the methods section. EELS maps for the individual chemical species (i.e., Te, Fe, Co, Ge) along with the ADF-STEM images are shown in Fig. S6 and Fig. S7 and exemplar individual EELS spectra are shown in Fig. S8. Fig. 2**a, b** shows the simultaneously acquired atomic resolution HAADF images for AA and AA' respectively, where we specifically focus on the Fe and Co EELS spectra. In order to distinguish between the atomic distribution of closely spaced (~1.3 Å apart) Fe (red) and Co (green) atoms, we calculated the intensity difference maps by subtracting the normalized Co intensity from the normalized Fe intensity as shown in the maps in Fig. 2**c,d**. The color scale used in the difference maps diverges from green to orange where green indicates enrichment of Co while orange indicates Fe enrichment. The zoomed-in HAADF and EELS maps that are displayed as insets in Fig. 2 show the location of these Co enrichment sites. In the AA phase, Co enrichment is mostly limited to the Fe-sites near the Ge atomic columns i.e., Fe2 sites. However, in the AA' phase there are two Co sites that exhibit enrichment, namely Fe2 which is close to the Ge atoms, and Fe1 which is close to the Te atoms.

To better grasp the fundamental origins underlying this topotactic phase transition between the centrosymmetric and polar phases of FCGT, density functional theory (DFT) calculations were performed using the exchange-correlation functional of Perdew, Burke, and Ernzerhof (PBE)[30] and pairwise van der Waals corrections[39] for the pristine stoichiometric endpoint compounds $Fe_5GeTe_2$ and $Co_5GeTe_2$, assuming collinear magnetic order and a fixed Fe1a site occupancy (see Methods for details). Specifically, the atomic positions and lattice parameters of $Fe_5GeTe_2$ and $Co_5GeTe_2$ for the ABC, AA, and AA' phases were relaxed for both ferromagnetic and antiferromagnetic inter-planar order until the Hellmann-Feynman forces on the atoms were below

0.02 eV/Å. The ground-state energy per formula unit (f.u.) was computed from each relaxed geometry and appears in Table S1, along with details of each configuration's lattice parameters.

The DFT ground-state energies of the structural and magnetic phases across all ABC and AA configurations lie within about ±10 meV/f.u., strongly supporting the hypothesis that each relevant polytype is experimentally accessible. Rather intriguingly, only ferromagnetic AA' $Fe_5GeTe_2$ exhibits an elevated ground-state energy by +94 meV/f.u. with respect to ferromagnetic ABC $Fe_5GeTe_2$, which indicates (as also confirmed by experiment) that the AA' non-centrosymmetric phase of FCGT is relatively more disfavored for pristine $Fe_5GeTe_2$. Yet this energetic difference is unobserved in $Co_5GeTe_2$ and is consistent with the fact that sufficient cobalt doping into $Fe_5GeTe_2$ may increasingly favor the AA' phase. From the relaxed lattice parameters in Table S1 we identify a significant reduction (1.7% for $Fe_5GeTe_2$ and 1.3% for $Co_5GeTe_2$) of the out-of-plane c lattice parameter in the AA' phase with respect to the AA and ABC structural phases, regardless of the transition metal species. This is consistent with the XRD measurements of the AA' FCGT crystal (Fig. S2 and Fig. S3), which show a reduction in c lattice parameter from AA to AA' phases of similar composition. The computed intralayer height across all relaxed structures/orders appear relatively similar in our DFT-PBE+D3 calculations, indicating the c lattice parameter reduction largely results from a change in interlayer spacing unique to the AA' phase. This reduction may also suggest that the distinct magnetic ordering behavior of this novel phase is influenced by a distance-sensitive interlayer exchange.

Temperature-dependent transport measurements of an AA' phase FCGT nanoflake exhibits a typical metallic behavior (Fig. 3**a**; inset shows the Hall bar device), and the corresponding magnetic field dependence of the Hall resistance ($R_{xy}$) at various temperatures is shown in Fig. 3**b**. At 10 K, $R_{xy}$ shows a rectangular hysteresis loop as a function of out-of-plane magnetic field, indicating a ferromagnetic ground state and a fully-remnant, single magnetic domain state. At ~ 70 K, the Hall data shows a sheared out-of-plane hysteresis loop suggesting the onset of labyrinthine domains.[31] At higher temperatures (70-320 K), the sheared hysteresis loop remains, while the saturation field gradually decreases. Finally, for temperatures higher than 325 K, the sheared hysteresis loops disappear. The value of $R_{AHE}$ is ~ 0.25 Ω at 10 K, and the saturated $R_{AHE}$ in Fig. 3**b** decreases with increasing temperature, revealing a surprisingly high $T_c$ of ~ 365 K (Fig. 3**c**), which is consistent with the bulk magnetization measurements (Fig. S10). The magnetotransport data along with the structural information in Fig.1 establish AA'-stacked FCGT as a polar, magnetic metal.

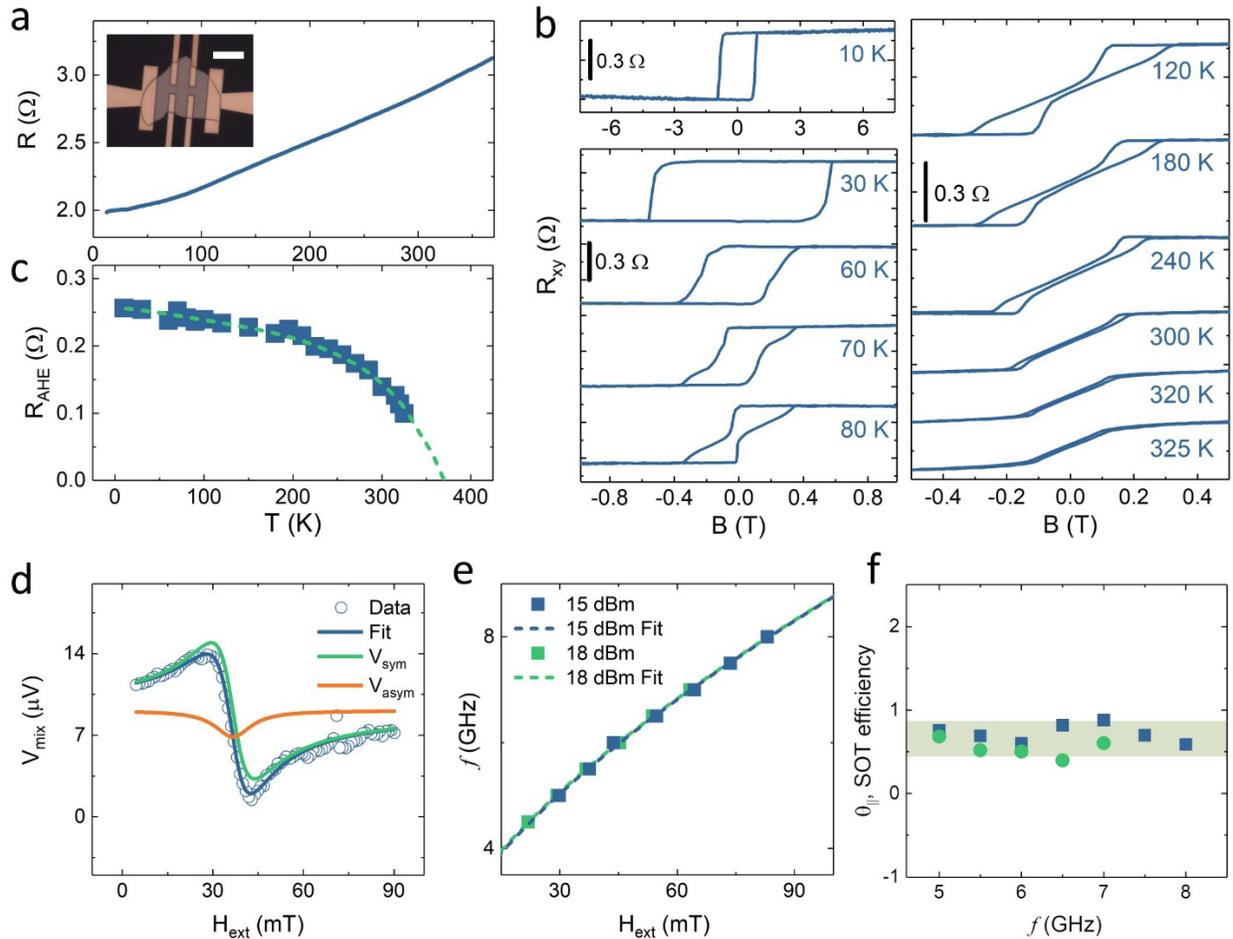

**Fig. 3: Magnetotransport and ST-FMR measurements of an AA' phase FCGT nanoflakes. a**, Temperature dependence of the longitudinal resistance. The Hall bar device shown in the inset of (a) is based on a 200-nm-thick FCGT nanoflake on a SiO$_2$/Si substrate. Scale bar is 10 μm. **b**, Hall resistance versus magnetic field at select temperatures. **c**, Temperature dependence of the anomalous Hall resistance in the saturation region. **d,** The ST-FMR signal of a FCGT (21 nm, ~11unit cells)/Co$_{0.9}$Fe$_{0.1}$(12 nm) sample at 5.5 GHz, 18 dBm. The solid lines are fits that show the symmetric (orange) and antisymmetric (yellow) Lorentzian contribution. **e,** ST-FMR frequency as a function of resonance field; the blue (15dBm) and green (18dBm) line is a fit to the Kittel formula. **f,** SOT efficiency (details in Methods section) as a function of frequency at 15 dBm and 18 dBm.

In wurtzite structure materials, due to the broken inversion symmetry, spin-orbit interaction causes splitting of the electronic band structure, leading to the Rashba effect. To verify current induced spin-orbit torques (SOTs) in the AA' phase, we performed Spin Torque Ferromagnetic Resonance (ST-FMR) measurements on multiple samples at room temperature. A typical ST-FMR signal for a FCGT/Co$_{0.9}$Fe$_{0.1}$ sample is shown in Fig. 3**d**; Fig. S11 shows details of the measurement. This spectrum can be fitted well to a sum of symmetric ($V_{sym}$) and asymmetric components ($V_{asym}$) (see Methods section), which are proportional to the in-plane damping-like, spin-orbit torque and out-of-plane torques, respectively. Fig. 3**e** shows the ST-FMR frequency as a function of resonant field, which is in good agreement with the Kittel formula, leading to an effective magnetization of the FCGT/Co$_{0.9}$Fe$_{0.1}$ bilayer of 626 kA/m. The SOT efficiency as the function of frequency is essentially constant, as shown in Fig. 3**f.** The average SOT efficiency ($\theta_\parallel$) is ~ 0.64±0.13, which is significantly larger than that of typical heavy metals[32].

We performed Lorentz TEM (LTEM) measurements on the AA' phase FCGT nanoflakes at room temperature to image the multidomain ground state suggested by the sheared hysteresis loops. LTEM imaging was performed on a 110 nm-thick FCGT flake, at a defocus of +4 mm at room temperature (Figs. 4**a**-4**c**). At zero applied magnetic field, no magnetic contrast was observed at normal incidence (Fig. 4**a**); the labyrinthine domains are visible (Fig. 4**b**) once the sample is tilted by 18° around the horizontal imaging direction as indicated in Fig. 4**a**. At zero tilt, the electron deflection cancels out due to the symmetric distribution of magnetization and thus the images exhibit no contrast in the standard LTEM mode.[33,34] An external field applied in the L-TEM along the beam direction triggers a domain morphology evolution from stripes to a mixture of bubbles and much shorter stripes (Fig. S12). When the magnetic field is increased up to 139 mT, we find that the bubbles clearly show dark/bright contrast perpendicular to the tilt axis, indicating Néel-type skyrmions (Fig. 4**c**). Finally, when the magnetic field is increased to 160 mT, the sample is magnetically saturated into a single domain state.

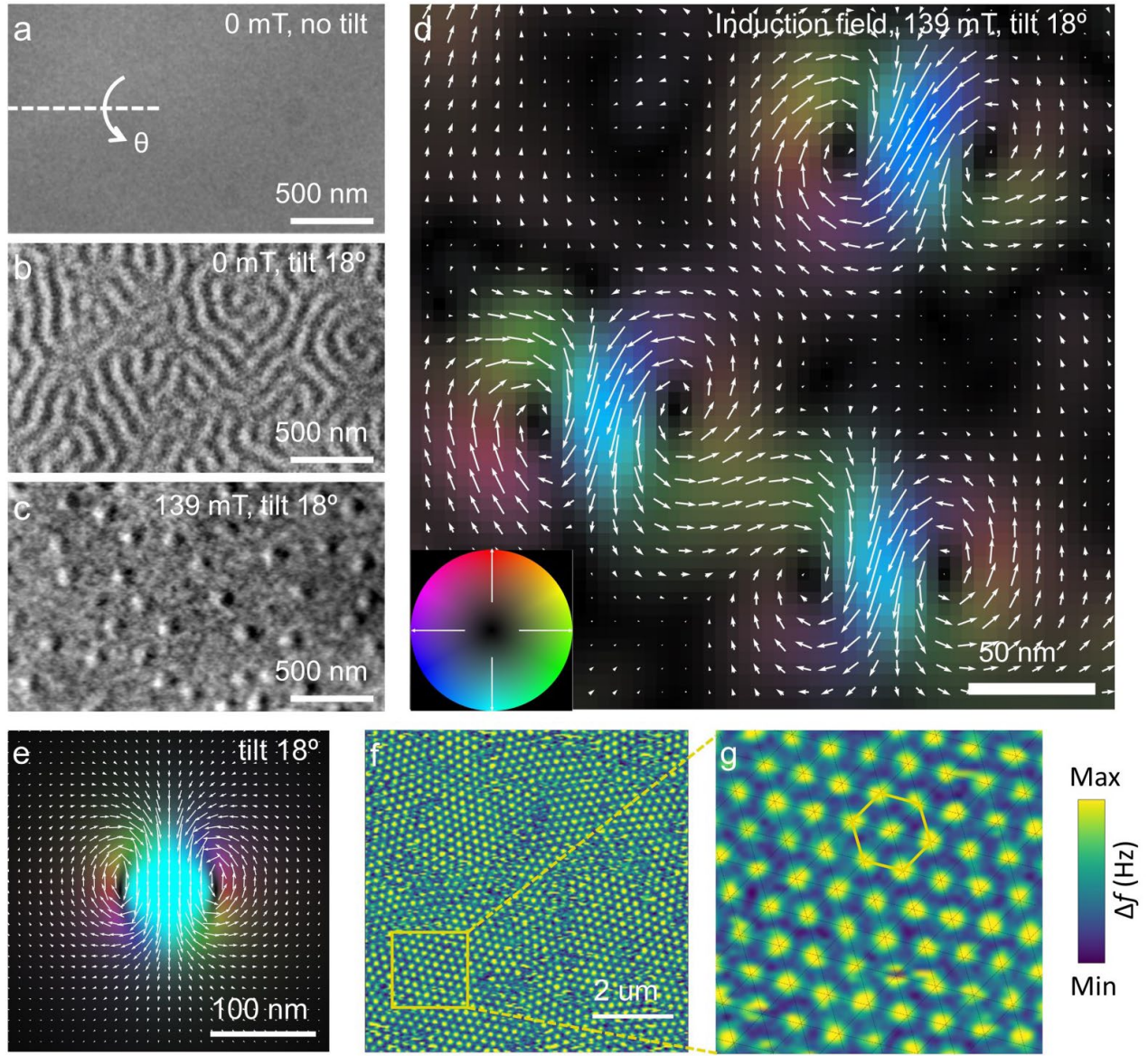

**Fig. 4: Room temperature spin texture of AA' phase FCGT. a, b,** Lorentz TEM images of the 110 nm-thick nanoflake acquired at the same region under zero-field and +4 mm defocus with 0°-tilt and 18°-tilt reveal a labyrinth phase with Néel character. **c,** In the same region at an 18°-tilt angle, isolated skyrmions emerge with an applied field of 139 mT. The skyrmion size is ~90 nm, determined from the lateral distance between the minimum and maximum intensity profile of the bubbles. **d,** Magnetic induction map obtained using four-dimensional Lorentz STEM (4D-LSTEM) equipped with an electron microscopy pixel array detector (EMPAD). The color and arrows indicate induction field components of Néel skyrmions perpendicular to the beam propagation direction for a 18° sample tilt. **e,** Simulated magnetic induction map for a Néel

skyrmion at an 18°-tilt. **f,** MFM images of a FCGT nanoflake on a SiO$_2$/Si substrate. The thickness is ~ 128 nm obtained by atomic force microscopy. **g,** High magnification of the skyrmion lattice region indicated by the yellow box in **f**. The skyrmion lattice parameter is ~ 200 nm corresponding to a density of ~23.3/μm$^2$.

To elucidate the detailed structure of such Néel-type skyrmions, we performed four-dimensional LSTEM (4D-LSTEM) experiments using a high-dynamic-range electron microscopy pixel array detector (EMPAD).[35] 4D-LSTEM works by acquiring a full two dimensional (2D), angle-resolved scattering distribution at each probe scanning position in a 2D grid, resulting in four-dimensional datasets. 4D-LSTEM works by detecting the phase shift of the electron beam induced by the lateral magnetic induction field of the spin textures (Fig. 4d). The magnetic induction field of Néel-type skyrmions (Fig. 4d) can thus be derived from the deflection of the electron beam in each diffraction pattern due to the Lorentz force, where the color and arrows represent the direction of the projected in-plane components.[36] A magnetic induction composed of clockwise and counter-clockwise spin curls was observed, which agrees well with the calculated induction field distribution for Néel-type magnetic skyrmions, Fig. 4**e** and Fig. S13.[34,37] The formation of skyrmions is also verified by magnetic force microscopy imaging illustrated in Fig.4(**f,g**). We observe an ordered hexagonal skyrmion lattice at room temperature.

The observation of a polar ferromagnetic metal at room temperature opens up several exciting fundamental opportunities, with the potential to impact applications in spintronics. Our work clearly demonstrates the role of structural and chemical order along with the subtle changes in electronic and spin structure through Co substitution triggers the changes in the broken inversion symmetry of the underlying crystal. This directly leads to the emergence of a bulk Dzyalozhinski-Moriya coupling and Rashba-type spin orbit coupling which are the fundamental requirements for both the formation of Néel -type skyrmions as well as the Rashba-Edelstein effect. This work shows enormous potential for crystal symmetry control of the spin and electric quantum state of the materials. Inspired by the magic-angle twisted bilayer graphene[38], spin-Moiré and skyrmion-Moiré patterns based on the Fe$_N$GeTe$_2$ system should exhibit novel spin textures and unconventional quantum states.

**Acknowledgements:**

H.Z., Y.T.S., J.R. and L.C. are supported by the Department of Defense, Air Force Office of Scientific Research under award FA9550-18-1-0480. X.C, S.S., E.B.C, R.B. and R.R. acknowledge support from the Quantum Materials program, funded by the U.S. Department of Energy, Basic Energy Sciences, Materials Sciences Division under contract DE-AC02-05CH11231. R.C. and J.Y. acknowledge the support by Intel Corporation under an award titled Valleytronics center. L.C. acknowledges financial support from the University of California Office of the President and the Ford Foundation. X.H. is supported by the SRC-ASCENT center which is part of the SRC-JUMP program. The electron microscopy studies were performed at the Cornell Center for Materials Research, a National Science Foundation (NSF) Materials Research Science and Engineering Centers program (DMR-1719875). The Cornell FEI Titan Themis 300 was acquired through NSF- MRI-1429155, with additional support from Cornell University, the Weill Institute and the Kavli Institute at Cornell. The authors thank M. Thomas, J. G. Grazul, M. Silvestry Ramos, K. Spoth for technical support and careful maintenance of the instruments. P.E. acknowledges funding by the Molecular Foundry, Lawrence Berkeley National Laboratory, which is supported by the U.S. Department of Energy under contract no. DE-AC02-05CH11231. The authors thank Rohan Dhall for FIB sample preparation. The devices were fabricated in the UC Berkeley Marvell Nanofabrication Laboratory.

**Author Contributions:**

H.Z. and R.R. designed the experiments. X.C. and E B.C. synthesized the single crystals. X.C. did the EDS and magnetization measurements. N.S.S. and X.C. did the single crystal diffraction measurement. R.C. and J. Z. performed nanoflake preparation, sample transfer and device fabrication. J. R. performed the ab initio calculations. H.Z. and X.H. performed the Hall and ST-FMR measurement. H.Z performed the AFM and MFM measurement. Y.T.S. and S.S. performed the STEM measurement with Z.C. and P.E. Y.T.S. performed the L(S)TEM measurements and simulation with Z.C. under supervision of D.A.M. H.Z., L.C., R.C. and R.R. wrote the manuscript. All authors discussed results and commented on the manuscript.

**Competing financial interests:**

The authors declare no competing financial interests.

**Data Availability Statement:**

The authors declare that all data supporting the findings of this study are available within the paper and its Methods sections. Raw data available from the corresponding authors upon reasonable request.

## Methods

### Sample synthesis

Single crystals of $(Fe_{1-y}Co_y)_{5-x}GeTe_2$ were grown by chemical vapor transfer method with iodine ($I_2$) as the transport agent. Starting materials comprised of elemental Fe powder (99.99%), Co powder (99.99%), Ge powder (99.999%) and Te powder (99.999%) with the nominal molar ratio 6(1-y):6y:1:5 were fully ground and mixed together inside the glovebox. About 50 mg of iodine was added to the mixture. The starting mixture was then vacuumed, back filled with 1/3 Argon and sealed inside the quartz tube with an inner diameter of 8 mm, an outer diameter of 12 mm and length of about 120 mm. The sealed quartz tube was placed horizontally inside a muffle furnace during the growth. The reaction temperature was set to 750 K under isothermal conditions for up to two weeks. Small and thin single crystals were harvested by quenching the furnace at 750 K in air. Excess iodine was removed from the surfaces of the crystals with ethanol.

### Single crystal XRD

Measurements for the $(Fe_{0.5}Co_{0.5})_{4.87}GeTe_2$ single crystal was performed on a Rigaku XtaLAB P200 equipped with a MicroMax 007HF rotating anode and a Pilatus 200K hybrid pixel array detector. Room temperature data were collected using Mo K$\alpha$ radiation ($\lambda$ = 0.71073 A). Data processing was done with CrysAlisPro and included a multi-scan absorption correction applied using the SCALE3 ABSPACK scaling algorithm within CrysAlisPro.[39] A solution was found using ShelXT in space group $P6_3mc$ (No. 186). An inversion twin law was applied to the structural model. Four sites were split 50:50 between Fe and Co (sites 1, 3-5) and one site was split 44:44 between Fe and Co (site 2), based on the freely refined occupancy observed at each site. This is consistent with the elemental composition observed by energy-dispersive X-ray spectroscopy (EDS). One tellurium atomic site occupancy was also allowed to freely refine the structure, resulting in an occupancy of ~94%. This was also consistent with that observed via EDS. The displacement parameters and atomic coordinates of the Fe and Co atoms were fixed to be equal using the EADP and EXYZ commands, respectively.

### DFT Calculations

All density functional theory calculations were performed with the VASP code, using the Perdew-Burke-Ernzerhof (PBE) exchange-correlation functional[30] with the electron-ion interaction treated with the projector-augmented wavefunctions (PAW) approach, accompanied by a pairwise Van der Waals correction via the zero damping DFT-D3 method of Grimme[39]. The plane-wave basis has an energy cutoff of 700 eV, self-consistent ground-state energy calculations use a convergence threshold of 0.01 meV and ionic relaxations use a convergence threshold of ±0.02 eV/Å for the Hellman-Feynman forces on the atoms. A Hubbard-U functional[46] of U = 0.3 eV is employed for all transition metal $d$-orbitals, and partial occupancies are identified using method of Methfessel-Paxton order-1 with a smearing parameter of σ = 0.04. The orbital occupancy smearing parameters and the Hubbard correctional term are tuned to address the itinerant nature of the $Fe_5GeTe_2$ and $Co_5GeTe_2$ systems. For this same reason, a sufficiently sampled k-grid is essential; a grid of 11×11×7 is used for 1x1x1 stoichiometric unit cells (ferromagnetic ABC phase) while a grid of 7×7×7 is used for all other ionic relaxation calculations consisting of 1x1x2 unit cells. In these simulations the Fe1/Co1 ions fixed to the 1a site. For pseudopotentials, Te ions used the valence configuration $5p^4 5s^2$, Ge ions used $4p^2 4s^2$, Fe ions used $3d^7 3p^6 4s^1 3s^2$, and Co ions $3d^8 3p^6 4s^1 3s^2$.

*Electrical and magnetic measurements*

High-quality FCGT nanoflakes were prepared on 100 nm $SiO_2$/Si substrates by means of mechanical exfoliation. E-beam lithography was utilized to pattern the samples, and then 3 nm Cr/150 nm Au was deposited with e-beam evaporation to fabricate Hall bar devices. The temperature-dependent resistance and Hall effect measurement were carried out in a CRYOGENIC measurement system (2 K, 14 T) with the magnetic field applied perpendicular to the nanoflake plane. A constant 10 μA *d.c.* current was applied for all the electric measurements. Magnetic characterizations were carried out with a superconducting quantum interference device magnetometer (Quantum Design, 2 K, 7 T), with the magnetic field applied along the out-of-plane and in-plane direction of the crystal.

*ST-FMR*

The test structure for ST-FMR measurements is comprised of $AlO_x$ (2.5 nm)/$Co_{0.9}Fe_{0.1}$(12.5 nm)/FCGT (21 nm)//$SiO_2$(100 nm)/Si substrate that was were patterned into rectangular bars by E-beam lithography. The ST-FMR measurements were carried out using a home-built system. The in-plane external magnetic field was oriented at 45° with respect to the current direction. During the measurement, a microwave current with fixed power and frequency was injected into the device and a DC voltage was collected while sweeping the external magnetic field. Multiple scans were accomplished at different frequencies to obtain the effective magnetization and spin-orbit torque efficiency. The ST-FMR signal can be fitted by: $V_{mix} = V_{sym}F_{sym} + V_{asym}F_{asym}$, where $V_{sym}$ and $V_{asym}$ are the amplitudes of the symmetric and antisymmetric components which are proportional to the in-plane damping-like torque and out-of-plane torques, respectively. The symmetric Lorentzian component is $F_{sym}$ and the antisymmetric Lorentzian component is $F_{asym}$,

$$F_{sym}(H_{ext}) = \Delta H^2/[(H_{ext} - H_r)^2 + \Delta H^2],$$
$$F_{asym}(H_{ext}) = \Delta H^2(H_{ext} - H_r)/[(H_{ext} - H_r)^2 + \Delta H^2]$$

where $\Delta H$ is the linewidth of the $V_{mix}$ curve, $H_{ext}$ is the external magnetic field, and $H_r$ is the resonance field. The effective magnetization ($M_{eff}$) of FCGT/Co$_{0.9}$Fe$_{0.1}$ bilayer is obtained by fitting the $H_r$ verse resonance frequency ($f$) with the Kittel formula:

$$f = \left(\frac{\gamma}{2\pi}\right)[(H_r + H_k) \times (H_r + H_k + 4\pi M_{eff})]^{0.5},$$

here, $\gamma$ is the gyromagnetic ratio, $H_k$ is in-plane anisotropy field.

The SOT efficiency ($\theta_\parallel$) is the ratio of the spin current density to RF current density,

$$\theta_\parallel = \left(\frac{V_{sym}}{V_{asym}}\right)\left(\frac{e\mu_0 M_s td}{\hbar}\right)[1 + (4\pi M_{eff}/H_{ext})]^{0.5}$$

here, $M_s$ is the saturation magnetization, $t$ is the thickness of the Co$_{0.9}$Fe$_{0.1}$ and $d$ it the thickness of FCGT.

## MFM

MFM experiments were performed using scanning probe microscopy (MFP-3D, Asylum Research) at zero magnetic field. The resonance frequency of the cantilever is ~ 75 kHz and the force constant is ~ 2.8 N/m. The MFM tip has a 40 nm cobalt alloy coating. The magnetic coating was magnetized perpendicular to the cantilever. During the MFM imaging, the distance between the tip and sample is maintained at a constant distance of 100 nm in a frequency-modulation mode. In order to prevent the surface oxidation of FCGT in air, a 5-nm Au capping layer was deposited over the FCGT flakes once the exfoliation is finished.

## Lorentz (S)TEM

A polypropylene carbonate (PPC)/Polydimethylsiloxane (PDMS) stamp was used to transfer the FCGT nanoflakes from SiO$_2$/Si substrates onto commercial TEM grids. Samples were transferred using a 3-axis stage to pick up FCGT from SiO$_2$/Si with PPC after a heating process at 90 °C and a subsequent cooling to 60 °C. Then, the FCGT was released onto TEM grids after the second heating to 90 °C. Finally, the residual PPC was dissolved in acetone for a clean surface of FCGT. We performed four-dimensional (4D) Lorentz scanning transmission electron microscopy (LSTEM) experiments using an electron microscopy pixel array detector (EMPAD), where the 2D electron diffraction pattern was recorded over a 2D grid of real space probe positions, resulting in 4D datasets. Experimental data was acquired using a FEI Titan operated at 300 keV with ~0.34 mrad semi-convergence angle, having a probe of ~3.5 nm FWHM (full-width at half-maximum). The LSTEM mode was realized with a new plugin which allows the continuous tuning of magnetic fields from 0 mT (residual field <2 mT) by adjusting the current of the objective lens. The magnetic field near the sample position was calibrated by using a custom-made Hall chip mounted on an electrical biasing holder (*Protochips inc.*). The magnetic deflection angle was determined by using a combination of edge detection and cross-correlation method with sub-pixel precision.[36]

### STEM

Cross-sectional TEM specimens were prepared on the same plan-view samples, using a FEI Strata 400 focused ion beam (FIB) with a final milling step of 2 keV to reduce damage. The initial sample surface was protected from ion-beam damage by depositing carbon and platinum layers prior to milling. The cross-sectional TEM specimen has a thickness of ~10 nm as determined by convergent beam electron diffraction analysis. HAADF-STEM images were recorded by using a Cs-corrected FEI Titan operated at 300 keV, with beam semi-convergence angle of 21.4 mrad and beam current of 20 pA.

### STEM simulation

The STEM simulations were carried out using the μSTEM software[40], with neutral atomic scattering factors of Waasmaier & Kirfel[41]. The atomic coordinates were taken from single crystal XRD refinements and density functional theory calculations for the AA' and AA stackings, respectively. The HAADF-STEM images of FCGT were simulated using a 300 keV electron probe with semi-convergence angle of 21.4 mrad, annular dark field collection angles of 70-200 mrad, and sample thickness of 70 nm. Here the simulations were carried out with a maximum scattering angle of 10.5 Å$^{-1}$ (204.1 mrad), and a real space sampling of 4.35 probe positions per Angstrom.

The thermal diffuse scattering effect was included with the frozen-phonon approximation, with 20 transmission functions and 150 Monte Carlo steps per probe position.

*EELS measurements*

Electron energy loss spectroscopy (EELS) mapping was performed using K3 Gatan direct electron detector installed at the back end of a Gatan continuum spectrometer in a TEAM I microscope at 300 kV. The probe angle, collection angle and current used for EELS acquisition were 20 mrad, 50 mrad and 100 pA respectively. The thickness of the samples determined from $t/\lambda$ ratio was around 10-20 nm for the AA and AA' configuration. The spectral EELS maps were generated by integrating the intensity of Fe, Co, Te and Ge after the power law background subtraction. The EELS maps and EELS line profiles were processed using Hyperspy.[42]

*Magnetic induction field simulation for Lorentz (S)TEM*

Magnetic induction field simulated from Bloch-type and Néel-type skyrmions were calculated from the magnetization distribution of an isolated skyrmion generated using the 360° domain wall model[43,44,45]. For simulation inputs, the diameter and domain wall width of skyrmions are set as 90 nm and 4.2 nm, respectively. The saturation magnetization $M_s$ is set as 301.6 kA/m, and sample thickness as 110 nm.

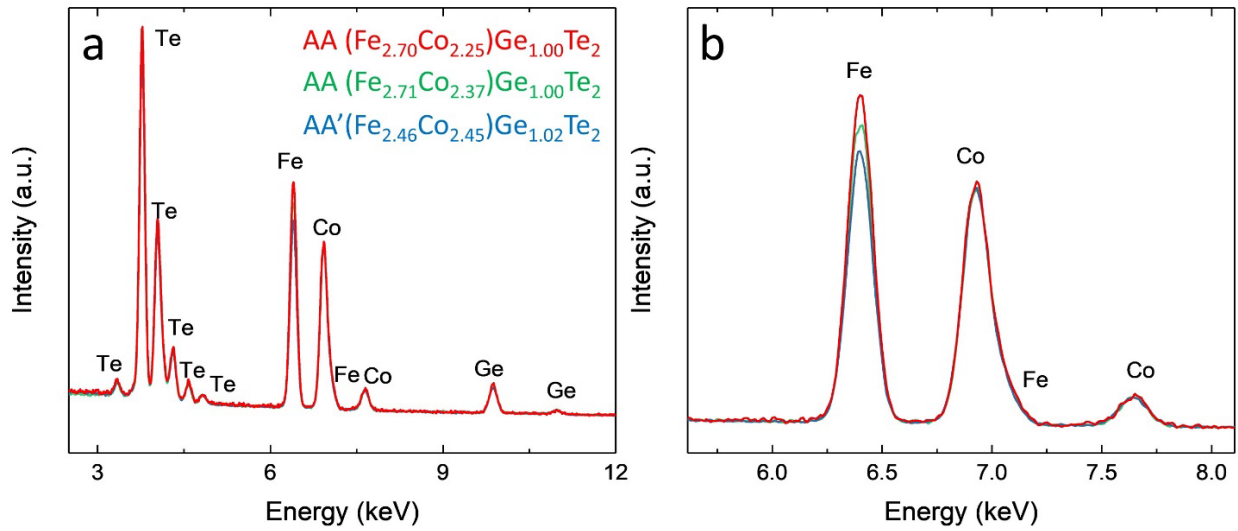

**Fig. S1 | Energy Dispersive X-ray Spectroscopy (EDS) of FCGT single crystal. a,** The EDS spectrum of AA (45% and 47%) and AA' (50%) phase FCGT single crystals. **b,** The zoomed-in spectrum near the Fe and Co peaks.

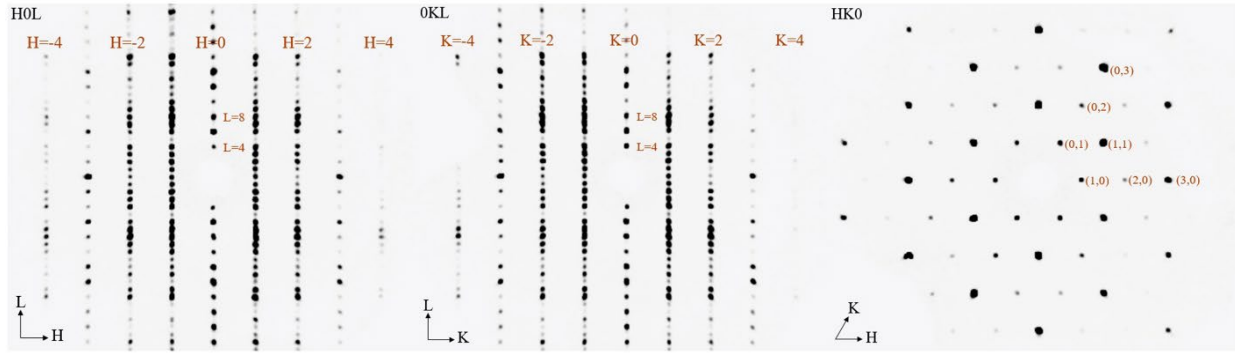

**Fig. S2 | Single crystal X-ray diffraction of AA' stacked FCGT.** X-ray diffraction data of FCGT single crystals indexed in the H0L, 0KL and HK0 planes. The FCGT crystals display a wurtzite structure with a space group P6$_3$mc from the refinement of scattering data. Wurtzite FCGT exhibits a non-centrosymmetric structure and is a hexagonal polar magnet metal which belongs to the $C_{6v}$ point group satisfying the prerequisite for hosting the Néel-type skyrmion and Rashba spin-orbit coupling.

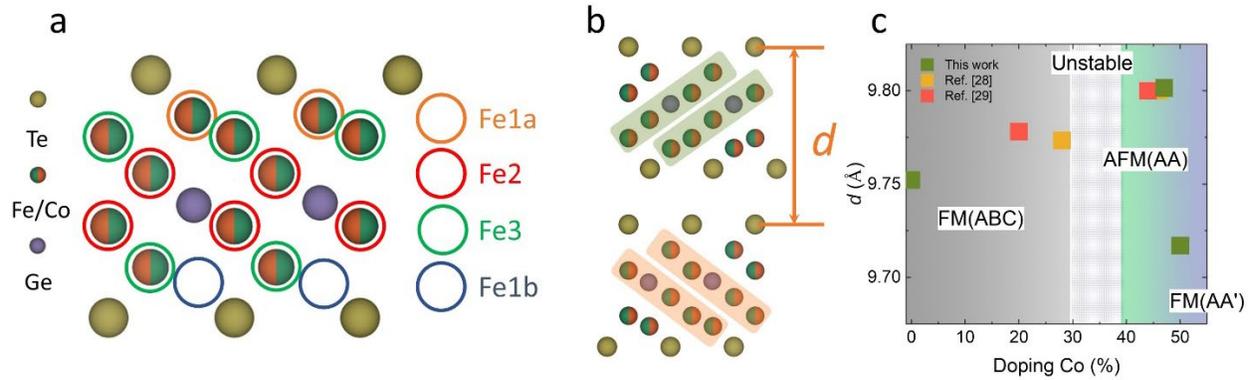

**Fig. S3 |Sublayer lattice constant along c axis with different cobalt concentrations. a,** Schematic of FCGT sublayer with atomic types. Fe/Co sites are labeled as Fe1a(orange), Fe2 (red), Fe3 (green) and Fe1b (blue). **b,** Side profile atomic rendering of FCGT for AA' stacking viewed along the [110] zone axis. **c**, As the cobalt concentration increases, the sublayer lattice constant correspondingly increases. At a cobalt concentration of 50%, the sublayer lattice constant drops abruptly due to the competition between vdW and Coulombic interaction. In the high cobalt doping region, the AA' phase can be stabilized.

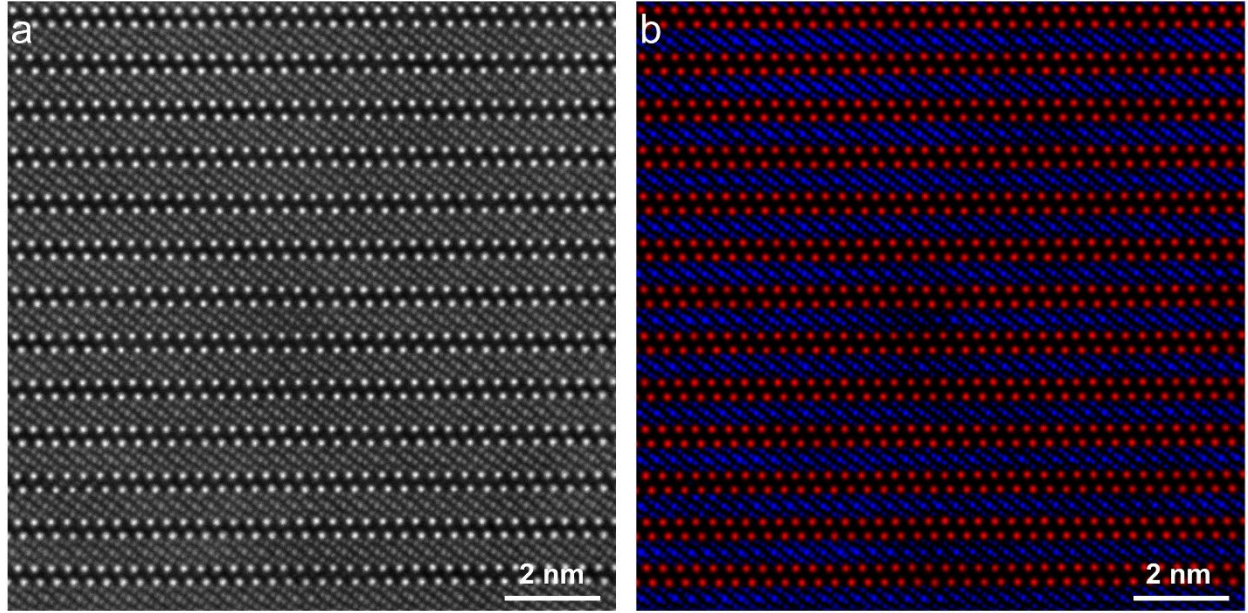

**Fig. S4 | Experimental HAADF-STEM image of AA FCGT (y=0.45) showing sublattices of Te and zag-FCG. a**, Raw experimental image. **b**, False colored sublattices of Te in red and zag-FCG in blue aided by 2D Gaussian fitting of atomic column positions.

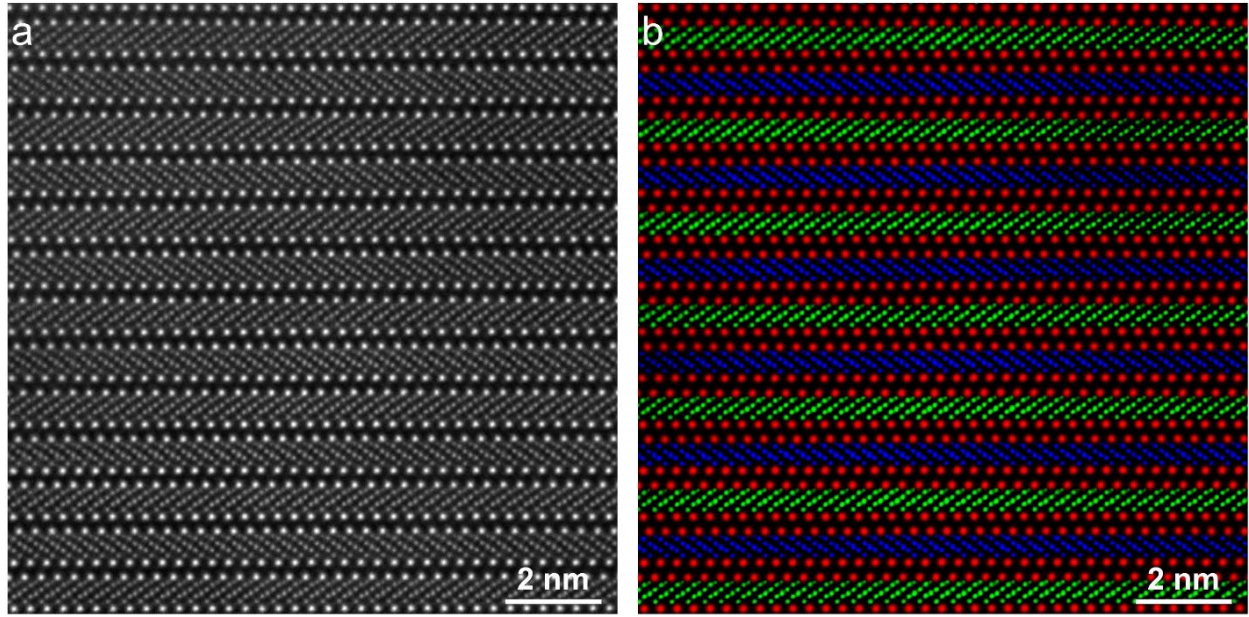

**Fig. S5 | Experimental HAADF-STEM image of AA' FCGT (y=0.50) showing sublattices of Te, zig-FCG, and zag-FCG. a**, Raw experimental image. **b**, False colored sublattices of Te in red, zig-FCG in green, and zag-FCG in blue aided by 2D Gaussian fitting of atomic column positions.

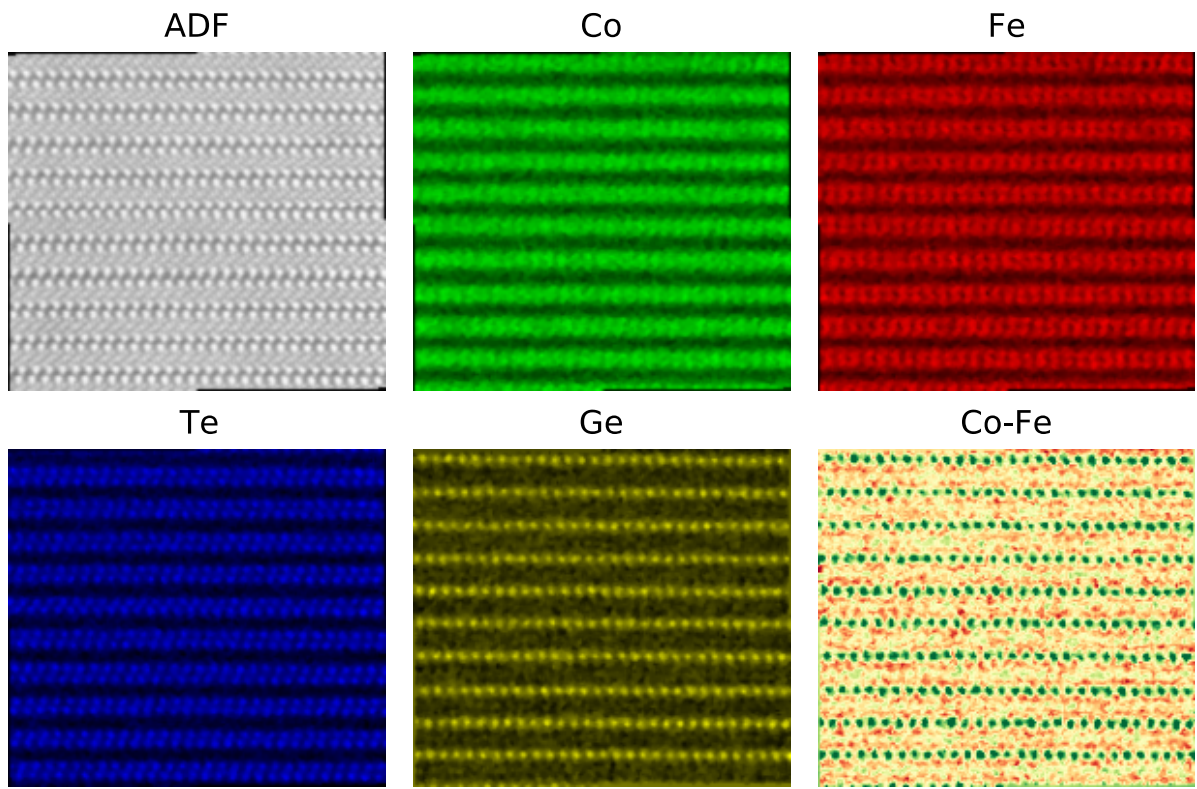

**Fig. S6| EELS mapping of AA phase FCGT. a**, Simultaneous HAADF, elemental EELS maps for **b** Co, **c**, Fe, **d**, Te and **e**, Ge. f, Normalized intensity difference maps (Co-Fe) for AA phase.

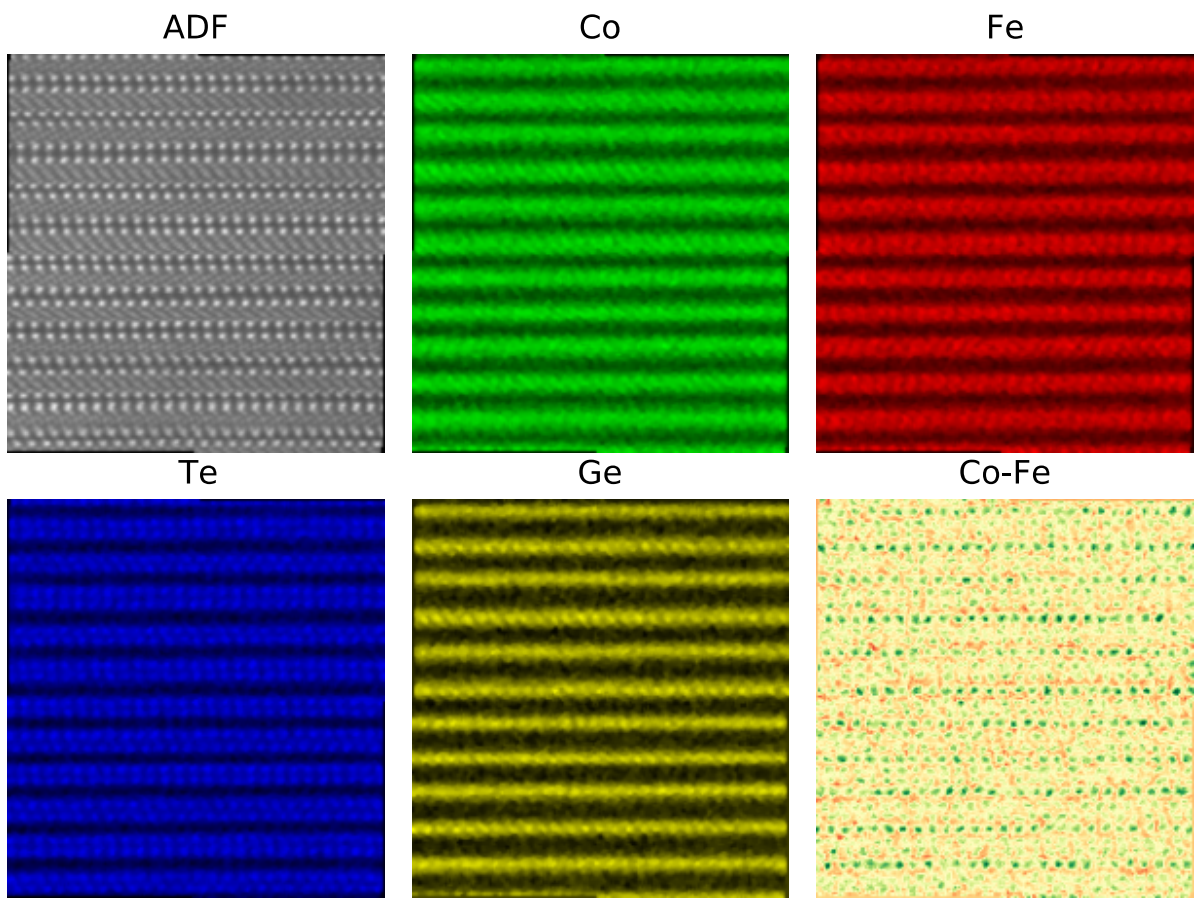

**Fig. S7| EELS mapping of AA' phase FCGT. a**, Simultaneous HAADF, elemental EELS maps for **b**, Co, **c**, Fe, **d**, Te, and **e**, Ge. f, Normalized intensity difference maps (Co-Fe) for AA' phase.

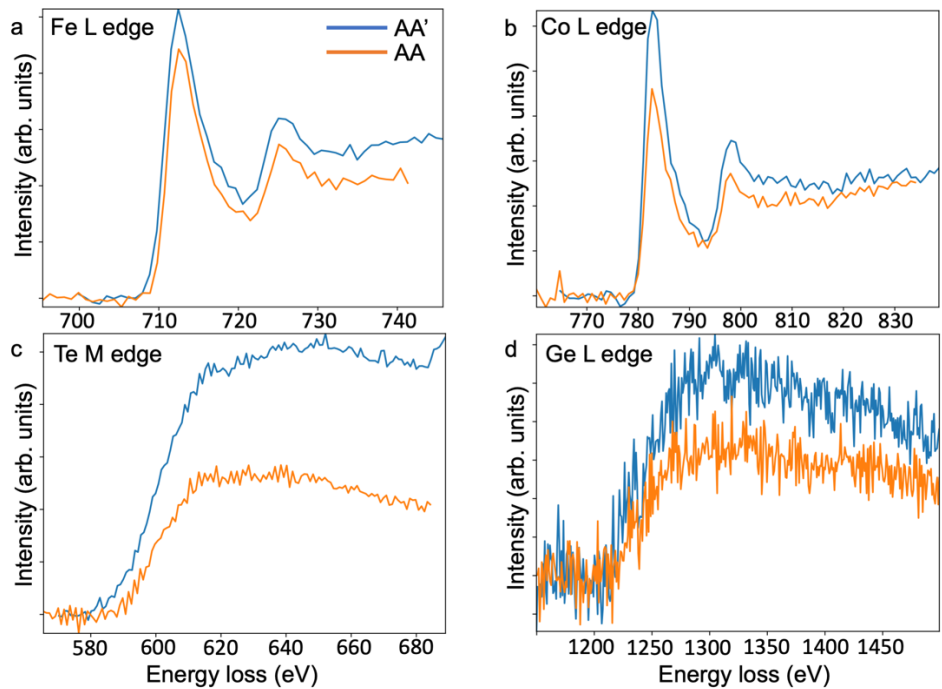

**Fig. S8| EELS spectral signals of AA and AA' phase FCGT. a,b,c,d,** Average EELS spectral signals from Fe L edge, Co L edge, Te M edge and Ge L edge for AA and AA' phase.

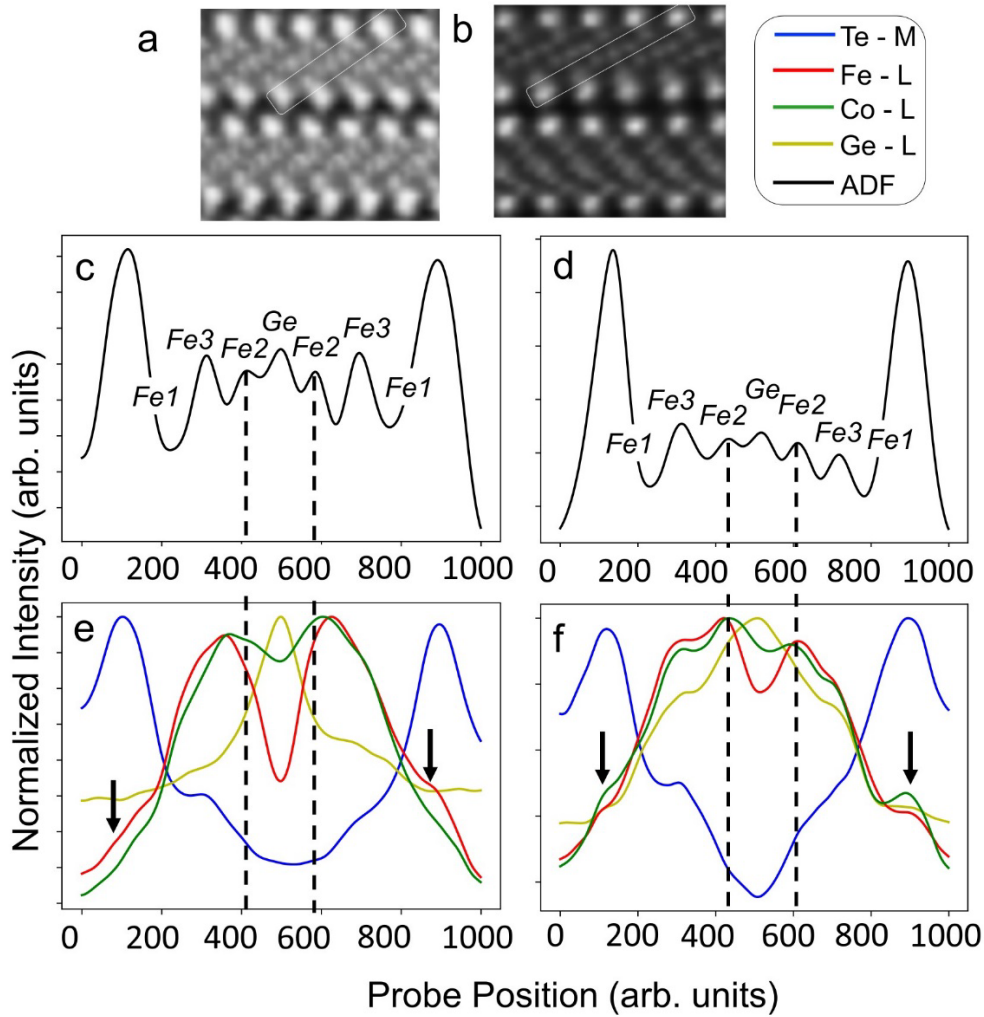

**Fig. S9| EELS Line profile.** Magnified ADF images of (**a**) AA phase and (**b**) AA' phase. (**c**) sum of ~50 EELS line profiles for Te-M, Fe-L, Co-L and Ge-L edge extracted along the atomic plane shown in ADF (**a, b**) for (**c**) AA and (**d**) AA' phase, respectively. Simultaneous summed line profile of ADF for (**e**) AA and (**f**) AA' phase. The legend for the line profile is presented beside (**b**). It is evident that there is Co enrichment in Fe2 sites in AA phase and AA' phase. However, at Fe1 site (indicated by the arrow), there is Fe enrichment in the AA phase and Co enrichment in the AA' phase.

| Structure | $E_{GS}$ (meV/f.u.) | $a$ | $c$ | Intralayer |
|---|---|---|---|---|
| $Fe_5GeTe_2$ **ABC (FM)** | — | — | — | — |
| $Fe_5GeTe_2$ **AA (AF)** | -4 | -0.0% | +0.1% | +0.4% |
| $Fe_5GeTe_2$ **AA' (FM)** | +94 | +0.1% | -1.7% | -0.3% |
| $Co_5GeTe_2$ **ABC (FM)** | — | — | — | — |
| $Co_5GeTe_2$ **AA (AF)** | -8 | +0.2% | +0.3% | +0.7% |
| $Co_5GeTe_2$ **AA' (FM)** | 0 | -0.2% | -1.3% | +0.4% |

**Table S1 | DFT-computed relative ground-state energies & crystallographic lattice parameters for myriad structural, magnetic, and chemical variations of the $(Fe_xCo_{1-x})_5GeTe_2$ compound.** Columns include the electronic ground-state energy (meV/f.u.), the in-plane and out-of-plane lattice parameters $a$ and $c$, and the intralayer spacing comprising a Van der Waals layer. All computed values are presented with respect to either the $Fe_5GeTe_2$ or $Co_5GeTe_2$ chemical endpoint compounds with ABC-phase and ferromagnetic order. The relaxed $Fe_5GeTe_2$ ABC (FM) structure has a ground-state energy of -51.3839 eV/f.u., along with $a$ = 3.975 Å and $c$ = 9.716 Å lattice parameters and $d$ = 6.641 Å intralayer spacing. Conversely, the relaxed $Co_5GeTe_2$ ABC (FM) structure has a ground-state energy of -45.1935 eV/f.u., along with $a$ = 3.975 Å and $c$ = 9.527 Å lattice parameters and $d$ = 6.539 Å intralayer spacing. Results with specific magnetic orders for each structural phase have been selected based on experimentally observed ordering.

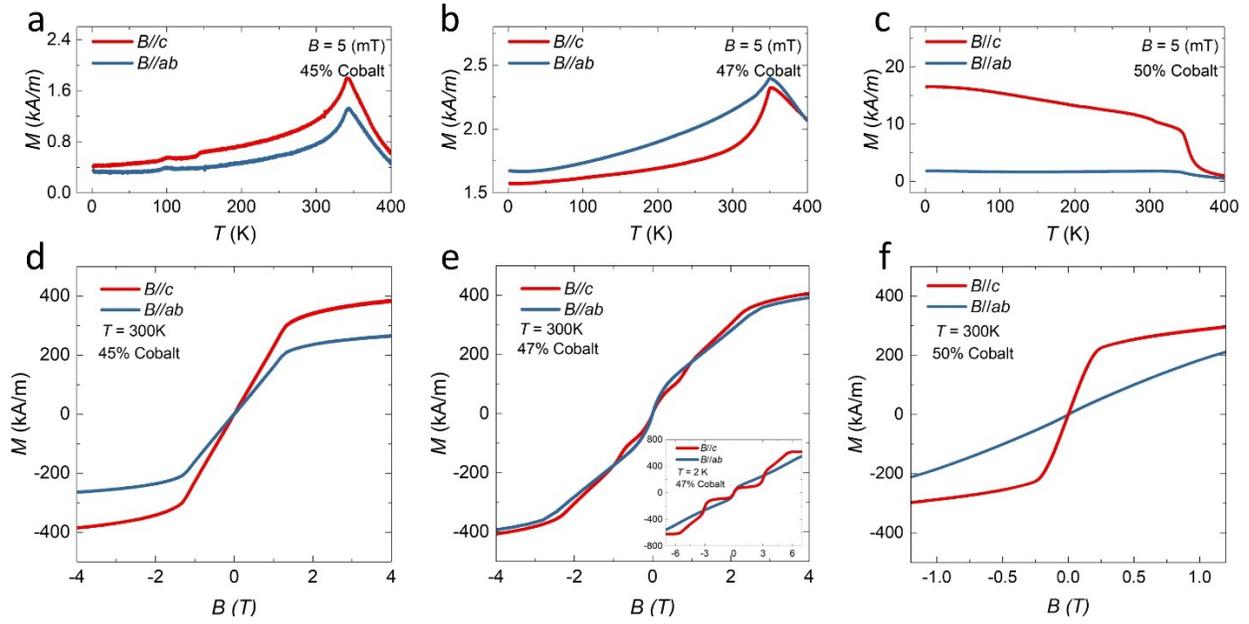

**Fig. S10 | Doping-concentration-dependent magnetic properties of FCGT single crystals.** Temperature-dependent magnetization of FCGT single crystals was obtained with a magnetic field of 5 mT. **a,b**, The magnetization of FCGT (45% and 47%) increases with increasing temperature and reaches a maximum at ~343K and 350 K, respectively, exhibiting an antiferromagnetic (AFM) to paramagnetic phase transition. **c**, The magnetization of FCGT (50%) decreases with increasing temperature and vanishes at ~365.5 K, showing a ferromagnetic (FM) to paramagnetic phase transition. Fig. **d**, **e**, **f** show isothermal magnetization curves of FCGT (45%, 47% and 50%) single crystals at room temperature. **d**, For a 45% cobalt concentration, a spin-flop transition is observed near 1.5 T. **e**, The magnetic field dependence of magnetization for a 47% cobalt concentration shows a step-like shape indicating a mixture of AFM and FM phases. Inset of **e** shows the isothermal M-H curves at 2K **f**, The isothermal M-H curves of the FCGT (50%) crystal exhibit an easy axis along the *c*-direction. A saturation magnetization, $M_S$ = 301.6 kA/m and an effective anisotropy, $K_{u,eff}$ = 2.4×10$^5$ J/m$^3$ are estimated.

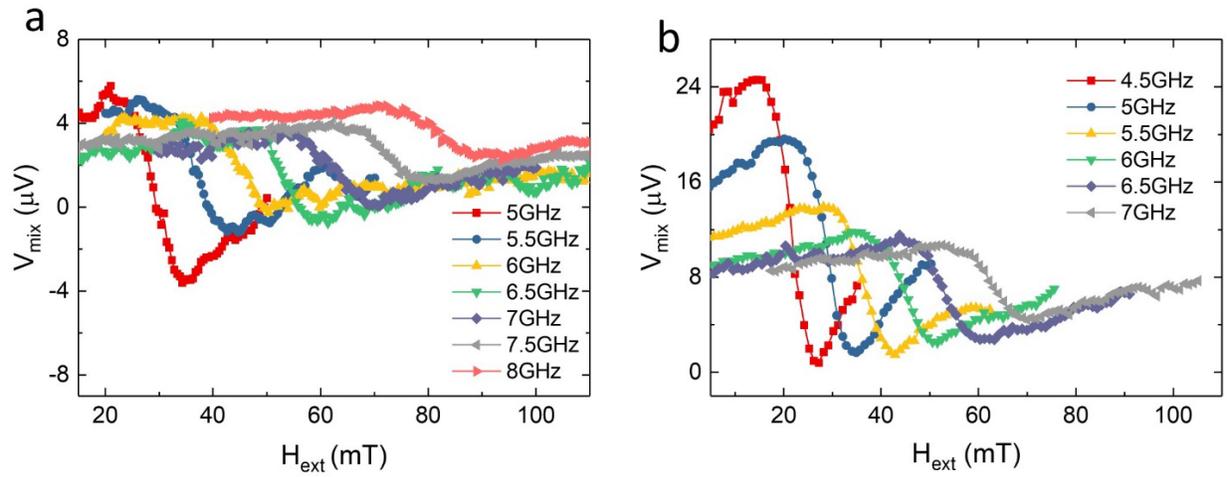

**Fig. S11| Frequency dependence of $V_{mix}$ for FCGT/Co$_{0.9}$Fe$_{0.1}$ bilayer. a, b,** Frequency dependence of $V_{mix}$ for FCGT/Co$_{0.9}$Fe$_{0.1}$ bilayer obtained at 15 dBm and 18dBm.

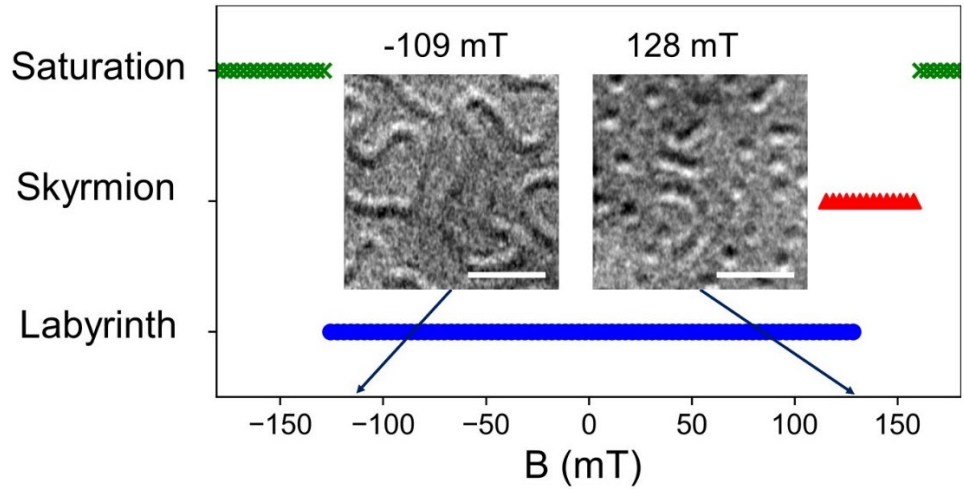

**Fig. S12| The evolution of spin texture with varying magnetic fields.** The phase diagram of magnetic texture versus applied magnetic field for a 110-nm-thick FCGT flake, with the external field swept from -180 mT to 180 mT at room temperature. Inset, two representative LTEM images acquired at -109 mT and 128 mT showing the serpentine domains and their mixture with skyrmions, respectively. It is clearly shown that the external magnetic field serves as an effective parameter to engineer the domain structures between a skyrmion (red triangle), saturated single domain (green cross), and labyrinthine (blue circle) phases. Scale bars in the insets are 500 nm. Lorentz TEM images were acquired at a defocus of +4 mm.

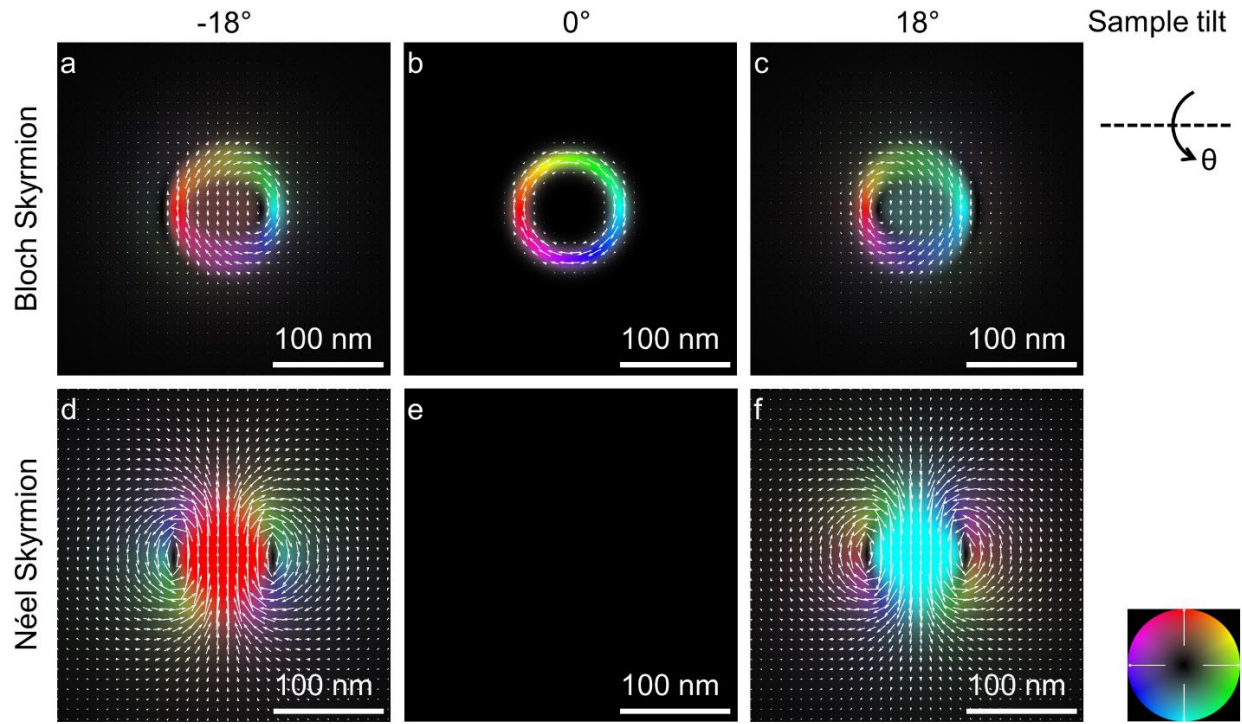

**Fig. S13| Simulated magnetic induction field from Bloch-type and Néel-type Skyrmions at different sample tilt angles.** Magnetic induction field distribution for a (**a-c**) Bloch skyrmion and a (**d-f**) Néel skyrmion at tilt angles of -18°, 0°, and -18° were calculated from the magnetization distribution of an isolated skyrmion generated using the 360° domain wall model. Since the phase gradient obtained from L-STEM is proportional to the induction field, the "biskyrmion" shape is a signature of a Néel-type skyrmion observed at a tilted angle. The simulation was carried out using a skyrmion diameter of 90 nm, and a domain wall width of 4.2 nm, a saturation magnetization $M_S$ of 301.6 kA/m, and a sample thickness of 110 nm. The slight shape difference between simulations and experiments may be explained by the distorted skyrmion shape (elongation) resulting from the experimentally applied external magnetic field.